\documentclass[referee]{cjaa}           
\usepackage{graphicx}                   
\input{epsf.sty}                        
\input{psfig.sty}                       

\begin{document}

   \title{Results of Recent Multi-wavelength Campaign of SS433
}

   \volnopage{Vol.0 (200x) No.0, 000--000}      
   \setcounter{page}{1}          

   \author{Sabyasachi Pal
      \inst{1,3}\mailto{}
   \and Sandip K. Chakrabarti
      \inst{1,2}
   \and K. Goswami
      \inst{3}
   \and A. Nandi
      \inst{2}
   \and B. G. Ananda Rao
      \inst{4}
   \and S. Mondal
      \inst{4} 
      }
   \offprints{Sabyasachi Pal}                   

   \institute{Centre for Space Physics, Chalantika 43, Garia Station Road, Kolkata 700084, India\\
             \email{space\_phys@vsnl.com}
        \and
             S. N. Bose National Centre for Basic Sciences, Salt Lake, Kolkata, 700098, India\\
        \and
             Jadavpur University, Kolkata 700032, India\\
	\and Physical Research Laboratory, Ahmedabad, India, 380009\\}

   \date{Received~~2004 July 25; accepted~~2004~~month day}

   \abstract{ We conducted multi wavelength campaign on SS433 in Sept. 
2002 using X-ray, B, Infra-Red and radio telescopes. We observed 
variabilities on a time-scale of few minutes in all the wavelengths.
We interpret them to be due to bullet-like features from
the accretion disk. We also present X-ray properties as obtained by RXTE.
   }

   \authorrunning{}            
   \titlerunning{}  

   \maketitle

\noindent To be Published in the proceedings of the 5th Microquasar Conference: 
Chinese Journal of Astronomy and Astrophysics


%
%
\section{Introduction}           
\label{sect:intro}
SS433 is a well known object which is ejecting matter in the form of two jets in  
symmetrically opposite directions at a speed of $v_{jet} \sim 0.26c$. 
Margon (1984) and Pal and Chakrabarti (2003) reviewed different theoretical and observational aspects of the source.
Grandi (1981) suggested that jets are not continuous and they are ejected from the compact 
object, like successive  
and discrete bullet-like entities, at least in the optical wavelength.
Since the bullets of energy $\sim 10^{35}$ ergs do not change
their speed for a considerable time ($\sim$ $1-2$ days), Chakrabarti et al. (2002) postulated 
that they must be ejected from accretion disk itself. They presented
a mechanism to produce quasi-regular bullets. Using results of
numerical simulations, they concluded that in the normal circumstances, 
a time interval of
$50$ - $1000$ s is expected in between successive bullet ejections.
These bullets are ejected from the X-ray emitting region and 
propagate through the optical, IR and radio emitting regions.
Thus, if the object is in a quiescence state, each individual bullet would be 
flaring and dying away in a few minutes timescale. This should be 
observable not only in optical wave length but also in all 
other wave lengths. So far, no such observations of individual
 bullets have been reported and it is necessary to make multi-wave length 
observation at relatively quieter states. In this paper, 
we present some results of our multi-wavelength studies.
 

\section{Observations}
\label{sect:Obs}
Radio observation was performed using Giant Meter Radio Telescope (GMRT) 
at $1.28$ GHz (bandwidth $16$ MHz). GMRT has $30$ antennas, each 
of $45$ meter diameter in nearly `Y' shaped array. Here we present 
results of 26th, 27th, 29th Sept. and 
1 Oct. 2002. 3C286 and 3C48 were used as flux calibrators and 2011-067, 
1925+211 and 1822-096 were used as phase calibrators. The data of the source 
is band-passed, self-calibrated and background subtracted.

\begin{figure}
\vspace{1cm}
   \begin{center}
\vskip -4.5cm
   \mbox{\epsfxsize=0.8\textwidth\epsfysize=0.6\textwidth\epsfbox{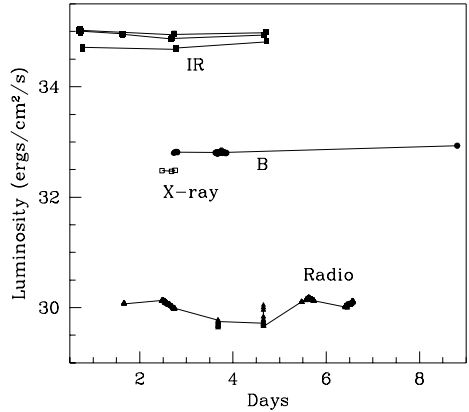}}
\vskip -1.0cm
   \caption{Multi-wavelength observation of SS433 from 25th Sept. 2002 to 1st 
Oct. 2002. Time is plotted in units of day ($0$ indicates 0:00 hour, 25th sept. 
2002 UT). The observations were made at Giant Meter Radio Telescope, Pune 
at $1.28$ GHz (Radio), State Observatory at Nainital (B band), $1.2$ m Mt. Abu 
Infra Red Telescope at $J$, $H$ and $K'$ bands and RXTE satellite 
($2-20$ kev) respectively. No correction due to extinction has been made.}
   \end{center}
\end{figure}

X ray observation was performed using the Proportional Counter Array 
(PCA) aboard RXTE satellite. 
We extracted the light curve from the XTE/PCA Science Data of Good
Xenon mode. We also extracted energy spectra from PCA {\bf standard 2}
data in the energy range $2$ - $20.0$ keV. For the spectrum, we
subtracted the background data.

\begin{figure}
   \begin{center}
\vspace{0 cm}
   \mbox{\epsfxsize=0.7\textwidth\epsfysize=0.7\textwidth\epsfbox{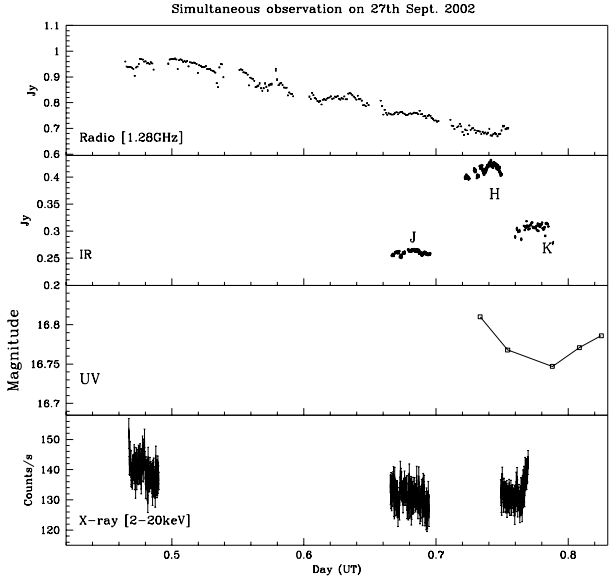}}
\vspace{-0cm}
   \caption{Multi-wavelength observation of short time variability in SS433 
by Radio (upper panel), Infra Red (second panel), B (third panel) and 
X-ray (lower panel) on 
27th September, 2002. The observations were made at Giant Meter Radio Telescope,
 Pune at 1.28 GHz (radio), 1.2 m Mt. Abu Infra Red Telescope at J H and $K'$ 
bands, Nainital State Observatory at B band and RXTE satellite ($2-20$ keV) 
respectively.}
   \end{center}
\end{figure}

IR observation was carried out using $1.2$ m Mt. 
Abu Infra Red Telescope equipped with Near - Infrared
Camera and Spectrograph (NICMOS) having $256$ $\times$ $256$ HgCdTe 
detector array cooled at $77$ K. We have used standard J ($\lambda$=1.25 $\mu$m, 
$\Delta\lambda$=$0.30$ $\mu$m), H ($\lambda$=$1.65$ $\mu$m, 
$\Delta\lambda$=$0.29$ $\mu$m) and $K'$ ($\lambda$=$2.12$ $\mu$m, 
$\Delta\lambda$=$0.36$ $\mu$m) bands. Observation took place
on 25th, 26th, 27th and 29th Sept. 2002 UT. J and H bands were binned 
at every $10$ second and $K'$ band were binned in every $20$ second. 
The data reduction were carried out using the IRAF package. All the 
object frames were de-binned, sky subtracted and flat fielded 
using normalized dome flats. GL748 was used as the standard star.   
  
Observation in B-band was carried out using Nainital state observatory
104 cm diameter telescope. We have observed on 27 and 28th Sept. 2002 UT 
using standard B band filter. The data was binned for every 20 minutes,
thus short time-scale variability could not be observed.

\begin{figure}
\vspace{-0cm}
{\hspace{-0cm}
\mbox{\epsfxsize=0.45\textwidth\epsfysize=0.45\textwidth\epsfbox{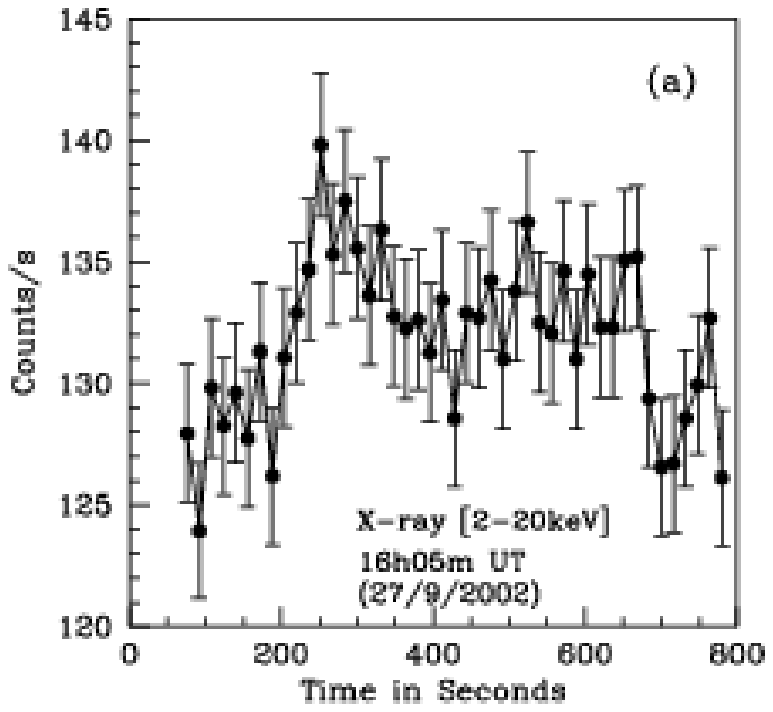}}}
\vspace{-0.3cm}
{\hspace{+0cm}   
\mbox{\epsfxsize=0.41\textwidth\epsfysize=0.41\textwidth\epsfbox{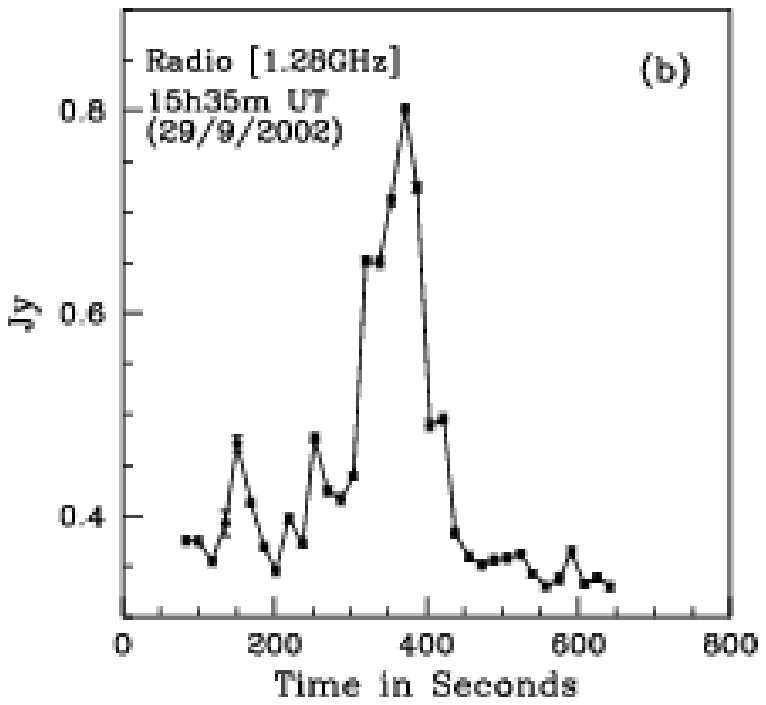}}}
\vspace{-0cm}
\caption{Individual flares in very short time-scale are caught. (a) An X-ray 
flare lasting $3.5$ minutes (observed on 27th Sept. 2002) and (b) a radio 
flare lasting $2.5$ minutes (observed on 29th Sept. 2002). Each bin size is 16 
seconds (from Chakrabarti et al. 2003).}
\end{figure}

\section{Results}
\label{sect:disc}
The observational results of multi-wavelength campaign from 25th Sept. 
2002 to 1st Oct. 2002 is shown in 
Fig. 1. Here, zero signifies 0:00 hour of 25th Sept. 
2002, UT. The results show a considerable variation in all the wavelengths. 
In Fig. 2, we have shown observational results of 27th Sept., 
2002. The first and second panels show the
radio and IR fluxes in Jansky, the third panel shows B band flux
in mag and the lower panel shows X ray counts
per second in $2$ - $20$ keV. 
Assuming isotropic emission, at a distance of $3$ kpc for the source, the
average radio, IR, X-ray and B luminosities are $1.1$ $\times$ $10^{30}$
 ergs/s, $5.5$ $\times$ $10^{34}$ ergs/s, $5\times10^{32}$ ergs/s and $10^{34}$ 
ergs/s respectively. Observations of radio, IR and B were carried out 
during 25th Sept. to 3rd October, 2002 and no signature of 
any persistent flare was observed. The radio
data showed a tendency to go down from $1.0$ Jy to $0.7$ Jy reaching at 
about $0.3$ Jy on 29th Sept., while the X- ray data showed a tendency to 
rise toward the end of the observation of the 27th. The IR data 
in each band remained virtually constant. The flux in H band was found 
to be higher than the flux in the J and $K'$ bands during 25-29th Sept. 2002. 
This could be possibly due to free-free emission in 
optically thin limit as discussed by Fuchs (2002).

From Fig. 2, we have seen that there are significant variation in time-scale of 
$T_{var}\sim 2 - 8$ minutes in all the wavelengths. In order to check 
if these are due to individual bullets, we present in Fig. 3a, 
one `micro-flare' like event from X ray data of 27th Sept. 2002. 
It shows significant brightening and falling in $\sim 100$ s.
The count rate goes up more than 15\% or so in about a minute. Similarly, 
in Fig. 3b, one micro-flare like event in radio is shown from 29th Sept. 
2002 data. Here the initial radio intensity (before the flare) was $\sim 0.3$ 
Jy, so that the micro-flare could be seen prominently. We observed brightening
the source from $0.35$ Jy to $0.80$ Jy in $\sim 75$ s which faded away in 
another $\sim$ $75$ s. That is, the intensity became more than double
in one minute! The energy contained in the radio micro-flare 
integrated over there lifetime is about $I\nu \tau 4 \pi D^2 10^{-23} =
1.1\times 10^{33}$ ergs (Here, I $\sim 0.8$ is the intensity in 
Jy, $\nu = 1280$ MHZ is the frequency of observation, $\tau \sim$ 100 s 
is the rise time of the bullet, D =$ 9$ $\times$ $10^{21}$ cm 
is the distance of SS433). Similarly, the energy contained 
in the X-ray micro-flare is about $\frac{1}{2}\tau (N_{\gamma,max}
-\bar{N}_{\gamma})E_{\gamma}4\pi D^2/A_{PCA}$ = $2.7 \times 10^{35}$ 
ergs (Here, $\tau \sim  100$ s is the rise time of the 
flare, $N_{\gamma,max}$ is the maximum photon count rate, $\bar{N}_{\gamma}
$ is the average photon number/s, $E_\gamma$ is the average photon 
energy, $A_{PCA}$ is the area of the PCA detectors).       
The spectroscopic study yields an average flux 
of $2.41$ $\times 10^{-10}$ ergs/c$m^2$/s. With an estimated 
duration of $100$ s, about 15\% energy is going to the 
micro-flare (Fig. 3a). The energy of this micro-flare 
is about $4.1 \times 10^{35}$ ergs. Since the 
radio luminosity is very small, even 
when integrated over $0.1$ to $10$ GHz radio band (with a 
spectral index of $\sim - 0.5$), we find that almost all the 
injected energy at X ray band is lost on the way during its
passage of $\sim 1 - 2$ d. 

\begin{figure}
\vspace{-3.9cm}
{\hspace {-0.5cm}
\mbox{\epsfxsize=0.9\textwidth\epsfysize=0.7\textwidth\epsfbox{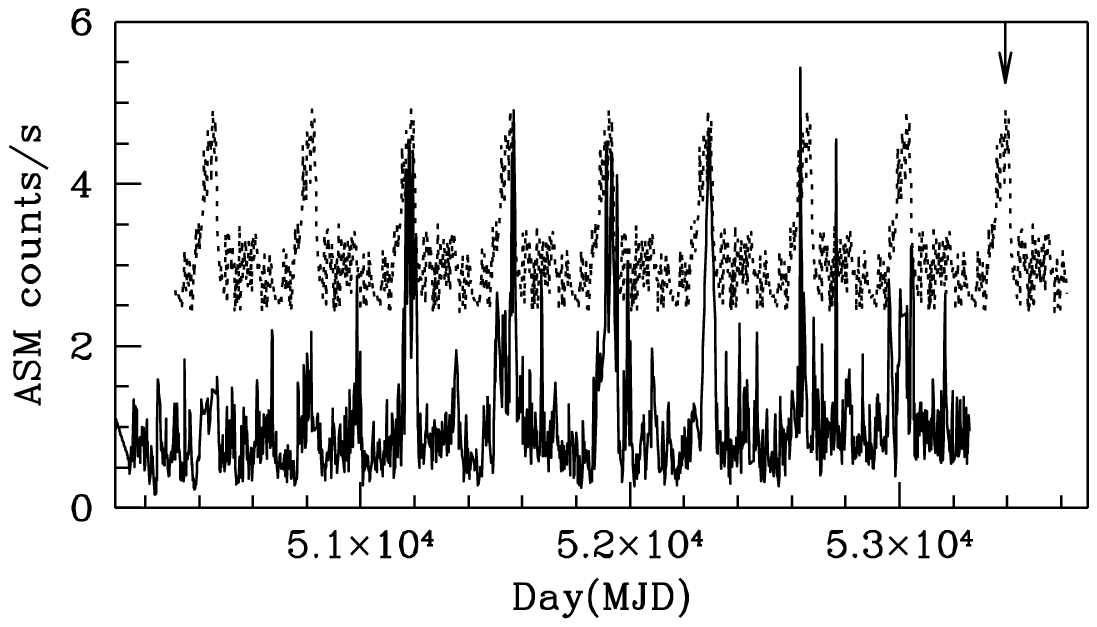}}}
{\vspace{-5.5cm}
\hspace {-6.5cm}
 \mbox{\epsfxsize=0.6\textwidth\epsfysize=0.7\textwidth\epsfbox{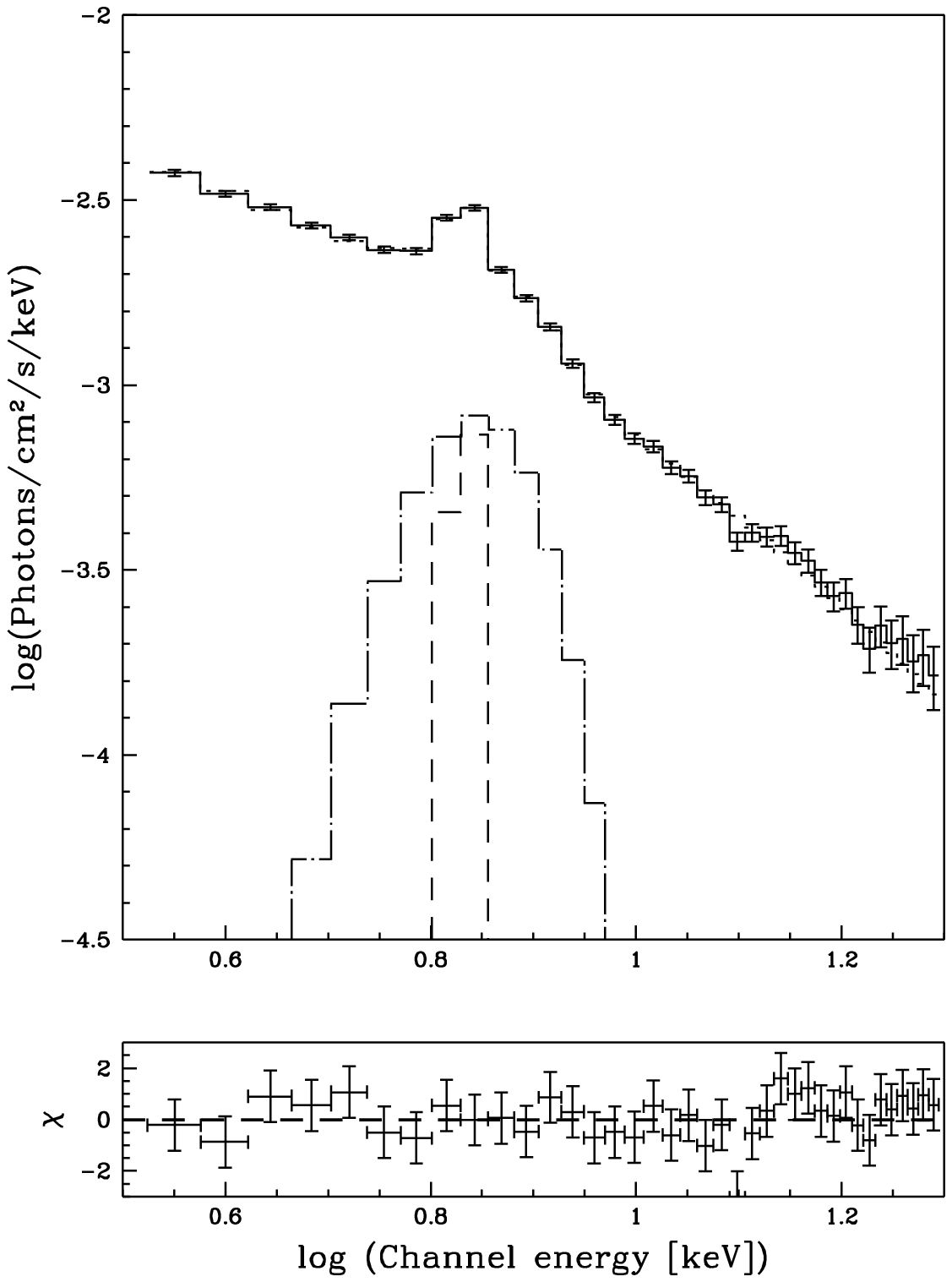}}}
\vspace{4.0 cm}
   \caption{Available RXTE/ASM lightcurve of SS433 (left) and X ray spectrum of SS433 on 27th September 2002 (right). The ASM is superposed by the folded lightcurve around 368d to show periodic
nature. Next predicted flare (peaking on Jan 22nd, 2005) is shown in arrow. 
In the lower right panel, the residuals are plotted (from Nandi et al. 2004.). }
\end{figure}
\begin{figure}
  \begin{center}
\vspace {-5cm}
   \mbox{\epsfxsize=0.9\textwidth\epsfysize=0.9\textwidth\epsfbox{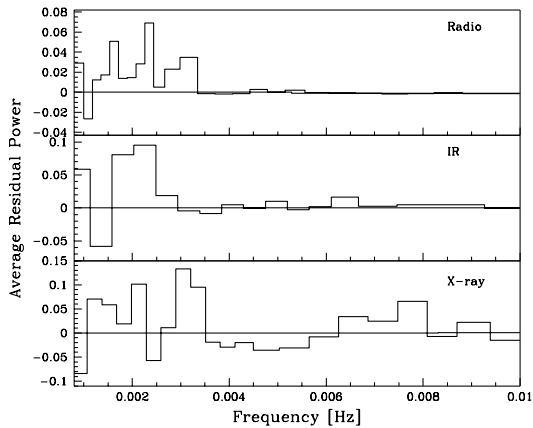}}
\vspace {-1cm}
   \caption{Residuals of the data obtained by
subtracting the suitable background noise from the Power Density 
Spectrum (PDS). Residuals of radio, IR and X-ray are shown in upper,
middle and lower panels respectively.}
  \end{center}
\end{figure}

In Fig. 4, left panel,  we present the available RXTE/ASM light curve of SS433.
We superpose on it (dashed) the lightcurve  folded around 368 days.
Note that flares are strongly periodic in nature with a periodicity of $368$ days (Nandi et al. 2004).
In the right panel, we  present the X-ray spectrum of SS433 
on 27th Sept. 2002 UT in which two iron lines were fitted. 
We took the Power Density Spectrum (PDS) of the variation in all the 
wavelengths in order to search for periodicities. We did not find any, but the
PDS does show a significant power in the frequency range of $\sim 0.002 - 0.008$ Hz. 
Deviation of the PDS from a power law
background $\propto \nu^{-\alpha}$ in all three bands gives an estimate
of excess power at low frequencies. 
The best fit values of $\alpha$ are $1.6$, $1.8$ and $1.9$ for radio, X-ray and IR, respectively.
Radio PDS shows excess at $\sim$ $0.0023$ Hz ($>3.2\sigma$), i.e. 
at $T_{var,r} \sim  7.2$ min and at $\sim 0.003$ Hz ($>1.6\sigma$), 
i.e. at $T_{var,r} \sim  5.5 $ min. X-ray power shows excess at $\sim$
$0.003$ Hz ($>2.7\sigma$), i.e. at $T_{var,x} \sim  5.5 $ min and 
at $0.0077$ Hz ($>1.4\sigma$), i.e. at $T_{var,x} \sim  2.1$ min. IR 
power shows excess at $0.0022$ Hz ($>4\sigma$), i.e. 
at $T_{var,ir} \sim 7.7$ min. These excesses (residuals) 
in PDS are shown in Fig. 5.

\section{Conclusion}
\label{sect:disc}

In this paper, we presented results of recent multi-wavelength observation in
X-ray, radio, B and IR wavelengths. We 
conclude that we may be observing ejection events of bullet-like 
features from the accretion disk in time-scales of 2-10 mins. 
Identification of small micro-flare events with those of bullet ejection
is derived from the time-scale of variabilities, which are roughly
the same in all the wavelengths. We find their presence in X-ray ($<
10^{11-12}$ cm ), IR ($< 10^{13-14}$ cm) and radio ($< 10^{15}$ cm)
emission regions.  We also find a periodicity in X-ray flaring bahaviour.

\acknowledgements We thank the staff of the GMRT who have helped us 
to make this observation possible. GMRT is run by the National Centre 
for Radio Astrophysics of the Tata Institute of Fundamental Research. 
We also thank Prof. R.  Sagar and Mr. Jeewan Pandey of Nainital Observatory 
for making the optical data available. This work is supported in part 
by a CSIR fellowship (SP) and a DST project (SKC and AN).

\label{lastpage}


\begin{thebibliography}{99}

\bibitem[]{}
Chakrabarti, S.K., Pal, S., Nandi, A., Ananda Rao, B.G., Mondal, S., 2003, ApJ, 595, L45
\bibitem[]{}
Chakrabarti, S.K., Goldoni, P., Wiita, P.J., Nandi, A., Das, S., 2002, ApJ, 576, L45
\bibitem[]{}
Grandi, S. 1981, Vistas Astr. 25, 7
\bibitem[]{}
Fuchs, Y. 2002, Ph.D. Thesis (SACLAY)
\bibitem[]{}
Morgan, B., 1984, ARA\&A, 22, 22, 507
\bibitem[]{}
Nandi, A.,Chakrabarti, S.K., Belloni, T., Goldini, P., 2004, MNRAS (submitted)
\bibitem[]{}
Pal, S. \& Chakrabarti, S. K. in `Recent trends in astro and plasma
 physics', 2003, (Eds.) S. K. Chakrabarti, S. Das, B. Basu \& M. Khan (CSP:Kolkata) 
\end{thebibliography}
\end{document}